*HYPOTHESIS:*

A possible role for stochastic astrophysical ionizing radiation events in the systematic disparity between molecular and fossil dates.

*by* ADRIAN L. MELOTT[1]
[1]Department of Physics and Astronomy; e-mail: melott@ku.edu
University of Kansas, Lawrence, KS 66045, U.S.A.

**Abstract:**
Major discrepancies have been noted for some time between fossil ages and molecular divergence dates for a variety of taxa. Recently, systematic trends within avian clades have been uncovered. The trends show that the disparity is much larger for mitochondrial DNA than for nuclear DNA; also that it is larger for crown fossil dates than stem fossil dates. It has been argued that this pattern is largely inconsistent with incompleteness of the fossil record as the principal driver of the disparity. A case is presented that given the expected mutations from a fluctuating background of astrophysical radiation from such sources as supernovae, the rate of molecular clocks is variable and should increase back to a few Ma, before returning to the long-term average rate. This is a possible explanation for the disparity. One test of this hypothesis is to look for an acceleration of molecular clocks 2 to 2.5 Ma due to one or more moderately nearby supernovae known to have happened at that time. Another is to look for reduced disparity in benthic organisms of the deep ocean. In addition, due to the importance of highly penetrating muon irradiation, the disparity should be magnified for megafauna.

**Running title:** Ionizing Radiation and Molecular Clocks

## 1. Introduction

There has been a long acknowledged disparity between ages determined from the fossil record and those derived from molecular divergence dating. (e.g. Hoareau 2014; Ho 2014; Pulquerio and Nichols 2007) The sign of this disparity is nearly always in the direction of much older molecular ages, but is reduced for very recent clades including domesticated animals (Ho and Larson 2006). A variety of effects may cause biases of either sign in the molecular ages, but, due to incompleteness of the fossil record, fossil ages are (aside from outright errors) lower bounds to the age of the taxon, so that the ages of originations are always underestimates. Since this could in principle account for most of the observed effect, much discussion has centered on this option.

A recent systematic study of trends within the disparities in avian clades (Ksepka et al. 2014) asserts that a variety of patterns, not to be enumerated here, are inconsistent with attributing all of the age disparity to gaps in the fossil record. We are urged to consider any possible systematic biases in molecular dates as well as calibration strategy. At this point it is relevant to mention that the whole basis of molecular dating has been accused of systematic underreporting of errors (e.g. Graur and Martin 2004). However, we shall for the purpose of this paper assume that they have a basis in reality.

One possible bias in molecular dates concerns a variable rate of molecular clocks due to changes in the mutation rate. An unknown but substantial fraction of mutations come from radiation of various kinds (Alpen 1997). The radiation background in the Earth's environment fluctuates strongly, so that it is expected to find events of increasing strength when looking further back in geologic time, since we are now not near a major energetic event (Erlykin and Wolfendale 2009; Melott and Thomas 2011). There is a normal operational assumption that molecular clocks move at a constant rate, but fluctuating radiation backgrounds would vary this rate. In the absence of selection pressures, isolated communities, etc. this variable mutation rate need not correspond to a correlated variable rate of evolution.

I wish to stress the following: this is a physics based hypothesis; there are many other biological explanations that may explain all or part of the phenomenon. I do not intend to claim that it is the best possible explanation, but only to introduce it for discussion.

High energy events are those which direct unusual amounts of ionizing radiation at the Earth. Examples are gamma-ray bursts and supernovae (Melott and Thomas 2011) as well as outbursts from the Sun often called solar proton events (Melott and Thomas 2012). A crucial characteristic is the amount of energy carried by individual protons, photons, or other constituents. If this energy is large, such as that carried by protons (cosmic rays) from a nearby supernova, it can produce air showers which have strong effects on the ground. In an air shower, the primary particle interacts with the atmosphere, and then produces a large number of other secondary particles in reactions with the atoms of the atmosphere. This shower cascades, ionizing the atmosphere which ultimately blocks most of the shower. Aside from the ionizing effect of the shower, only muons reach the ground in sufficient abundance to have direct radiation effects on organisms living there or in the ocean. Since such very strong events are not occurring now (see e.g Overholt et al. 2015), molecular clock rates determined from very recent data would not include this acceleration. The purpose of this note is to suggest further examination of this possibility, with attention to possible tests.

Ksepka et al. (2014) reported that the disparity between fossil and molecular clocks is greater with mitochondrial DNA than with nuclear DNA. This observation is consistent with our hypothesis of a radiation link for the disparity, because mitochondrial DNA is more subject to damage from radiation and to oxidative stress, one of the primary mechanisms of radiation damage to DNA (Yakes and Van Houten 1997; Kam and Bonati 2013).

In what follows, we give more detail on recent advances in knowing the history of the radiation background near the Earth, and some of the characteristics of the radiation that reached the surface. Then, we suggest some expectations, which show ways that the hypothesis of radiation-coupled mutation might be tested.

## 2. Radiation Events and the Earth

There are a variety of possible types of astrophysical radiation and likely sources for events at the Earth (Melott and Thomas 2011). Dominant among these are the Sun (see Wdowczyk and Wolfendale 1977) and other stars in our galaxy. It has been known for some time that supernovae and gamma-ray bursts from other stars in our galaxy are likely, based on their intensity and frequency of occurrence, causal agents in mass extinction every few 100 Myr (Melott and Thomas 2011 and references therein). The Sun has X-ray flares, but the dominant form of solar radiation for biological consideration is in Solar Proton Events. There was in 775 AD an event indicated by $^{14}$C in tree rings which exceeds anything in the modern era (Miyake et al. 2012; Jull et al. 2014) and which is probably attributable to the Sun (Melott and Thomas 2012; Usoskin et al. 2013). All of these are potentially dangerous sources (e.g. Thomas et al. 2013), although interpreting the new data on the Sun and Sunlike stars is an emerging area.

The atmosphere provides considerable shielding, but effects on the ground can still potentially include radiation in the form of muons (Atri and Melott 2011; Marinho et al. 2014) and neutrons (Overholt et al. 2013; Overholt et al., 2015). Neutrons rarely penetrate below the stratosphere. Most muons are stopped by a kilometer of water, so any potential muon damage would not include benthic organisms in the deep ocean. In addition, the ionizing radiation can deplete the stratospheric ozone layer (Thomas et al.

2013, and references therein), admitting increased damaging ultraviolet-B (UVB) from the Sun. Nearby supernovae will bombard the Earth with much higher energy cosmic rays (nuclei of atoms, mostly protons) than are likely to come from the Sun.

The primary mechanism of radiation damage to the biosphere emphasized to date is UVB from a depleted ozone layer (Melott and Thomas 2011). All of the events described can cause this to varying degree. This UVB may be damaging to plankton and induce skin cancer, but it is unlikely to penetrate deeply enough inside metazoans to cause germ line mutation. Most forms of primary and secondary radiation do not penetrate in great amounts lower than the stratosphere—where they can damage the ozone layer. Muons from most events studied are too few in number to significantly add to the radiation background on the ground (Overholt et al. 2015).

Recent results (Thomas et al. 2016) change the picture. Computing the effects of an indicated (Melott 2016 and references therein) supernova at 100 pc from the Earth, we found that (for an event inside the Local Bubble, a region of hot gas in which the Earth resides), most effects were negligible—except for the muons. The reason is that the very high energy cosmic rays from a supernova far exceed the intensity found at the Earth from any other type of event that has been considered. We found that a 20-fold increase in muons on the ground was indicated for thousands of years from a single burst event—effectively tripling the local radiation level. Muons are very different from most other sources of ionizing radiation. They interact very weakly, so that they penetrate all living things and even up to a kilometer of water. On the other hand, they are so abundant that in spite of very little interaction, they comprise a significant level of background radiation. Most kinds of ionizing radiation will not penetrate far into living organisms, but muons will pass right through. For this reason, they will reach inside even the largest animals. Effects of muons will be a nearly unique signature of nearby supernovae. In the local supernova case (Thomas et al. 2016), other radiation effects were relatively small. The amount of time the irradiation lasts can be increased if the cosmic rays generating the muons are trapped with the Earth inside a structure such as the Local Bubble.

It is worth mentioning that the other novel effect associated with nearby supernovae is ionization of the troposphere (Thomas et al. 2016) with effects as yet not well understood. However, this is unlikely to be related to the topic at hand.

Newer estimates suggest that the events associated with the $^{60}$Fe detections are more likely to have been at 50 pc than at 100 (Fry et al. 2016; Mamajek 2016). We have work in progress to more fully describe the effects of such events, which will certainly be more intense than at 100 pc.

### 3. Reasonableness of the Idea

Most astrophysical radiation including the indirectly increased UVB can be stopped by 10 m of water, which would exclude many benthic organisms. However, DNA of such organisms may be affected in the pelagic larval stage for those larvae that spend most of their time near the ocean surface. Organisms which are shielded from the radiation should not show effects of strong fluctuations. In particular, deep-water benthic organisms should display more congruence between the fossil and molecular dating methods. It now appears (Thomas et al. 2016) that although ozone depletion/UVB enhancement is important for a variety of events, the episodic nearby supernovae at 50-100 pc will have their dominant impact through muons. Muons will affect everything except organisms below about 1 km of water (or hundreds of meters of rock). So, UVB effects stop at about 10m depth and muon effects at about 1 km depth.

Kspeka et al. (2014) noted that for very old dates (many times 10 Myr), there is better congruence between molecular and fossil estimates than for younger dates. This would be consistent with a high rate of mutation in the last few Myr, and a return to a geologic mean rate over longer periods. These trends are also suggested by Ho and Larson (2006), who put the high mutation rate at 1-2 Ma, and also suggest that the mutation rate has been low in the last few 10's of kyr. The combined picture is then of a lull at the present, a peak at the last few Myr, and a return to moderate rates over 10's of Myr.

We note that an increased mutation rate may facilitate, but does not necessarily imply a speeding up of evolution. There are many other important effects, such as isolation of populations, changes in environmental factors, etc. which are equally important, such as the closing of the Isthmus of Panama near the P-P boundary.

There is a large amount of new data in the form of $^{60}$Fe in sediments which suggest (Fry et al. 2015, 2016; Melott 2016; Wallner et al. 2016; Breitschwerdt et al. 2016) that one or two supernovae went off within one or a few hundred light years of the Earth around the beginning of the Pleistocene. Breitschwerdt et al. (2016) use numerical modeling to suggest that as many as 14 more went off in the previous 8 Myr, giving rise to the Local Bubble as a kind of blast region. The isotope data is found in ocean sediments (Wallner et al. 2016); lunar rock samples (Fimani et al. 2016); cosmic rays collected directly in space (Binns et al 2016), and fossil magnetotactic bacteria (Ludwig et al. 2016). This data comprises a dramatic confirmation and extension of earlier terrestrial samples (Knie et al. 1999, 2004). This would indicate an enhanced radiation environment persisting for at least several thousands of years for each event. Therefore, if this idea has merit, there should be an acceleration of the molecular clocks relative to the fossil record around 2.5 Ma. The transport modeling work which examined the distribution of the isotopes through the interstellar medium, and the formation of the Local Bubble by the blast waves from the supernovae (Breitschwerdt et al. 2016) was published simultaneously and done without knowledge of the new data from the other studies. Improved modeling is in progress which should better constrain the dates.

In the meantime, our modeling work suggests increases in ionizing radiation on the surface and in the upper ocean persisting for at least thousands of years for each supernova, and longer if we are trapped in the Local Bubble with it. We note that the suggested timing for the increased mutation rate (Ho and Larson 2006) is roughly coincident with the times estimated for the nearby supernovae.

## 4. Testing the hypothesis

We have argued that this suggestion appears reasonable based on existing data on isotope distribution at the Earth, times of higher mutation rate, and simulations of the radiation which may penetrate to the ground from the supernovae which generated the isotopes. We will now try to go beyond the reasonableness argument based on existing data, and make some predictions which we would expect to be true based on the hypothesis that astrophysical ionizing radiation events episodically increase the mutation rate in the molecular clock.

(1) Even muons will not penetrate much beyond a km of water. We therefore expect major reduction in the disparity between molecular and fossil clocks for deep-sea organisms.
(2) Very large organisms, which were common during the Pliocene and Pleistocene, would still be easily penetrated by muons, and not by other forms of ionizing radiation (except for inhaled radon, not expected to be a significant result of supernovae). Therefore, the 20-fold or larger increase in muons indicated by supernova events in the last few Myr would be more significant for these organisms, and we would see a concomitant significant increase in the clock disparity for them around the P-P boundary.
(3) Mitochondrial DNA, which shows an increased mutation rate, should show a disparity concentrated around the P-P boundary.
(4) Improved data on molecular clock rates and times of increased radioisotope deposition on Earth from supernovae should continue to coincide beyond the crude up-then-down rate trend presently observed. There may be a second maximum around 7 Ma (e.g. Breitschwerdt et al. 2016).

In closing, it should be emphasized that we do not claim that this is better than other existing explanations for this disparity, but that it is a possibility that should be considered, and can be tested by looking for the effects listed above.

**Acknowledgements:**

I am grateful to Larry D. Martin (deceased), who introduced me to this problem several years ago.  Helpful comments have come from three anonymous referees, Drew Overholt, Richard Bambach, Bruce Lieberman, Arnold Wolfendale, and Greg Edgecombe. This research was supported by NASA Exobiology and Evolutionary Biology grant NNX14AK22G.

There are no competing financial interests.